\documentclass[aps,prl,preprint,groupedaddress]{revtex4-2}

\usepackage{graphicx}

\begin{document}


\title{Spatial-spectral mapping for long-duration broadband terahertz pulse generation in on-chip waveguide arrays}


\author{Yibo Huang$^1$}
\author{Yao Lu$^1$}
\email[]{yaolu@nankai.edu.cn}
\author{Haoyu Duan$^1$}
\author{Chao Wang$^1$}
\author{Xitan Xu$^1$}
\author{Jiwei Qi$^1$}
\author{Qiang Wu$^{1, 2}$}
\email[]{wuqiang@nankai.edu.cn}
\author{Jingjun Xu$^{1, 2}$}
\email[]{jjxu@nankai.edu.cn}
\affiliation{$^1$ The Key Laboratory of Weak-Light Nonlinear Photonics, Ministry of Education, TEDA Applied Physics Institute and School of Physics, Nankai University, Tianjin 300457, China}
\affiliation{$^2$ Collaborative Innovation Center of Extreme Optics, Shanxi University, Taiyuan, Shanxi 030006, China}


\date{\today}

\begin{abstract}
Conventional approaches to terahertz (THz) pulse generation are restricted by the Fourier-transform limit, which hinders the creation of sources that combine long duration with broad bandwidth—a capability crucial for many spectroscopic and sensing applications. In this work, we overcome this challenge in the terahertz domain using an on-chip gradient waveguide array. The key is to spectrally disperse the pulse into spatially separated channels within a lithium niobate chip, effectively decoupling the design of temporal and spectral properties. We validate the source by distinguishing amino acid mixtures, demonstrating its tailored biosensing potential. This work establishes a novel mechanism for integrated THz generation, offering considerable promise for broadband spectroscopy and on-chip photonics.
\end{abstract}


\maketitle

\section{introdution}
Governed by the Fourier transform principle, the inverse relationship between pulse duration ($\Delta$$\tau$) and spectral bandwidth ($\Delta$$\omega$) constrains light-matter interactions to either narrowband continuous-wave or ultrashort pulse regimes \cite{Weiner20081028}. As a result, the generation of long-duration, broadband pulses capable of particular light-matter interaction remains elusive. Chirping, using dispersion to temporally stretch a pulse, provides a long duration but not implicating a solution, since it staggers frequencies in time without prolonging the matter-interaction time for any single frequency component \cite{STRICKLAND1985447,Jonusas:24,jolly2019spectral,10081337,Eggert:18,10.1063/1.111543,OKeeffe:12}.

To circumvent this limitation, gradient metasurfaces have been proposed \cite{Ding_2018,doi:10.1126/science.1210713}. As shown in Fig. 1(a), they achieve a broadband response by mapping different frequency components to distinct “artificial atoms” across the surface. Unlike chirped dispersion, which separates frequencies in time, this approach decouples them in space. This spatial-spectral mapping theoretically allows each frequency component to interact with matter independently and simultaneously, thereby preserving both broad bandwidth and a long interaction window. This concept has led to versatile photonic devices for wavefront shaping \cite{doi:10.1126/science.aax2357,zhao2013tailoring,ni2013ultra,li2013spin,aydin2011broadband,Pors:15,doi:10.1126/science.1098999}, nonlinear optics \cite{https://doi.org/10.1002/adma.202307494}, advanced imaging \cite{doi:10.1126/science.aag2472,doi:10.1126/sciadv.1601102}, and highly sensitive biosensing \cite{https://doi.org/10.1002/adma.202314279,aigner2024continuous,doi:10.1126/science.aas9768,doi:10.1126/sciadv.aaw2871}.

However, diffraction imposes a fundamental constraint when such spatially encoded pulses propagate in free space \cite{D2NR01561G}, especially under broadband pulsed illumination \cite{https://doi.org/10.1002/adma.202208947}. Theoretically, once the separated frequency components radiate from the metasurface, their wavefronts expand and undergo spatiotemporal overlap in the far field. This overlap induces frequency-dependent interference, which disturbs the precise spatial-spectral mapping and undermines the engineered decoupling of temporal and spectral properties. Consequently, the simultaneous long-duration and broadband characteristics are degraded upon propagation, reinstating the conventional trade-off that the spatial design sought to overcome.

\begin{figure*}[tp] 
    \centering 
    \includegraphics[width=0.95\textwidth]{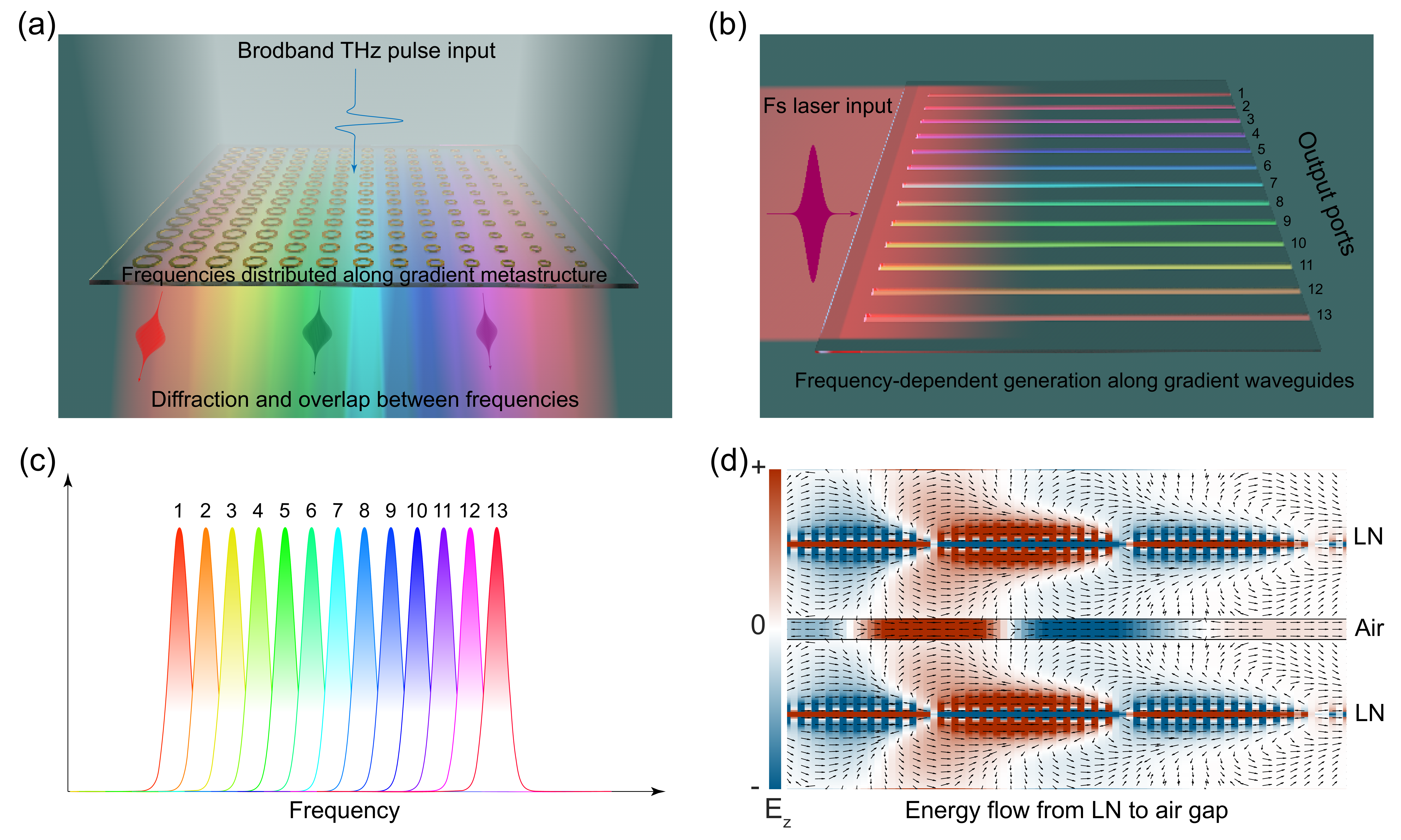}
    \caption{ Generation of long-duration broadband THz pulses. (a) High-quality broadband response using gradient metasurfaces. (b) Generation of long-duration, broad THz pulses using a specially designed on-chip gradient waveguide array. (c) The narrowband THz waves generated in different channels forms a long-pulse, broadband THz pulse. (d) THz generation process: the THz waves are generated within LN pillars, parts are coupled and confined to the low-loss air gap while propagating.} 
    \label{FIG12} 
\end{figure*}

In this Letter, we overcome this diffraction limit by demonstrating the generation of long-duration, broadband terahertz (THz) pulses on a monolithic chip. Our approach utilizes a specially designed gradient waveguide array in lithium niobate, as illustrated in Fig. 1(b). When femtosecond (Fs) laser pulses are laterally focused into a lithium niobate subwavelength waveguide (LNSW), the frequency of the generated THz wave is determined by the local effective permittivity. By engineering an array of such waveguides with gradually varying properties, we distribute different THz frequency components into distinct, diffraction-free spatial channels within the chip. This on-chip spatial-spectral mapping preserves the engineered decoupling of frequencies, effectively preventing the spatiotemporal overlap and interference that degrade performance in free space [Fig. 1(b)]. The coherent synthesis from all waveguide ports yields a long-duration, broadband THz pulse ideal for light-matter interaction studies [Fig. 1(c)]. Notably, a portion of the generated THz wave is coupled into and confined by a low-loss air gap within the chip structure during propagation, enhancing transmission efficiency \cite{Almeida:04,Xu:04} as shown in Fig. 1(d). To validate the principle, we theoretically demonstrate the capability of this tailored source to identify biomolecular mixtures. Crucially, this is accomplished without mechanical frequency or delay scanning \cite{koch2023terahertz,liebermeister2021optoelectronic}, presenting a simplified and robust conceptual platform for integrated broadband spectroscopy.
\section{results}
To achieve controllable generation of narrowband THz waves at a target frequency, we employ a slot waveguide array (SWA). The proposed SWA design is illustrated in Fig. 2(a). It consists of periodically alternating lithium niobate (LN) pillars and air slots, with designed widths l and s, respectively, oriented along the LN crystal axis to maximize the optical nonlinearity for efficient THz generation. Figure 2(b) shows the fundamental-mode electric field distribution of the THz wave within the SWA. Owing to boundary conditions \cite{Almeida:04,Xu:04}, the field is strongly confined within the air slots, which can enhance light-matter interactions \cite{yang2009optical}. The propagation characteristics are governed by three geometric parameters: the width l and thickness h of the LN pillars, and the width s of the air slots. Figure 2(c) presents the calculated dispersion relation for a designed SWA ($l$ = 133 $\mu$m, $h$ = 50 $\mu$m, $s$  =  8 $\mu$m), obtained using finite-difference time domain (FDTD) method, along with that of the pump laser pulses (white dotted line). Under the lateral excitation configuration \cite{Yang:17,https://doi.org/10.1002/lpor.202000591}, the phase-velocity matching condition is satisfied at the intersection of the two dispersion curves, corresponding to a frequency of 0.36 THz. This selective velocity matching enables the efficient generation of narrowband THz radiation at this specific frequency. As shown in Fig. 2(d), the generated THz waveform evolves from a single-cycle to a multi-cycle oscillation as it co-propagates with the pump pulse along the waveguide.

\begin{figure}[htbp] 
    \centering 
    \includegraphics[width=\textwidth]{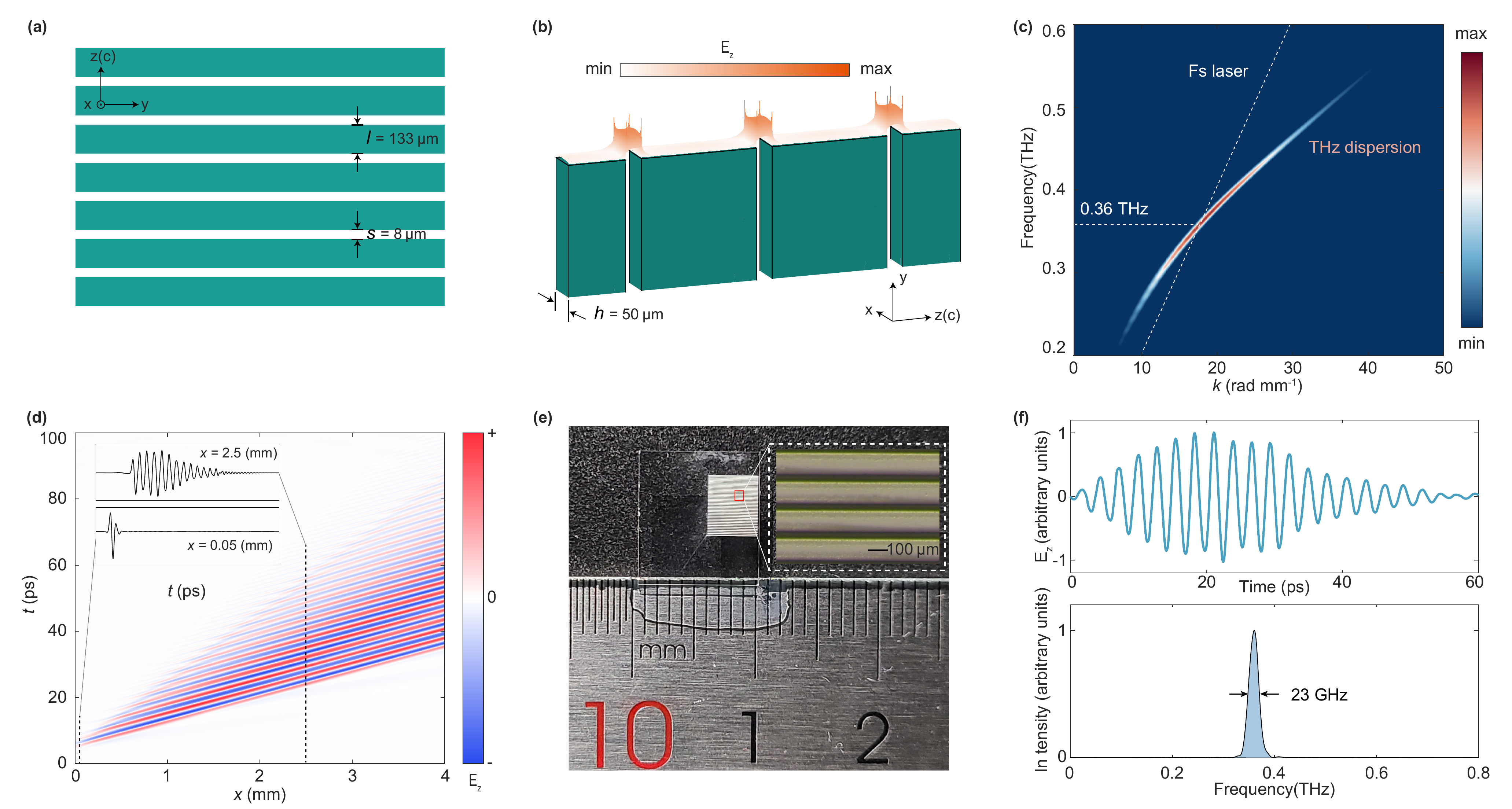}
    \caption{ The narrowband THz generation of a specific frequency. (a), The schematic of the proposed SWA. (b), The electric field distribution of the THz wave within the SWA. (c), The dispersion relations of the THz wave and the femtosecond laser pulse. The gray dotted line represents the pump laser pulses, which has a constant effective group index of 2.264. (d), The simulated time and space traces of THz E-field in the SWA. The inset depicts time traces of the THz E-field generated at two positions: $x$ = 0.05 mm and $x$ = 2.5 mm, showing that the oscillations increase as propagating forward. (e) Appearance of the fabricated SWA with designed geometric parameters. (f), Time traces of the THz E-field generated at the detection point in the fabricated SWA (top panel). The spectrum of the generated THz pulse in the fabricated SWA (bottom panel).} 
    \label{FIG2} 
\end{figure}

The designed SWA was fabricated on a 50 $\mu$m-thick $x$-cut MgO:LN chip using femtosecond laser direct writing \cite{sivarajah2013chemically} (Supplementary Note 1). An optical micrograph of the fabricated sample is shown in Fig. 2(e). The overall chip measures 10 mm in width and 11 mm in height, with the 4-mm-wide SWA positioned adjacent to one edge. For THz generation and detection, a femtosecond laser-based electro-optic sampling system was employed \cite{leitenstorfer1999detectors} (Supplementary Note 2).

Theoretically, the duration of the generated THz pulses increases with propagation distance under lateral excitation. Practically, however, the finite length of the SWA (4 mm) and the gradual dissipation of the pump beam limit the achievable pulse length. The measured THz pulse, detected at a point outside the SWA, lasts approximately 60 ps; its temporal waveform is shown in the top panel of Fig. 2(f). Fourier transformation of this signal yields the corresponding spectrum (bottom panel, Fig. 2(f)), which exhibits a central frequency of 0.360 THz and a bandwidth of 23 GHz, in good agreement with the dispersion-based prediction.

By tailoring the geometric parameters of the SWA to engineer its dispersion relation, the center frequency of the generated narrowband terahertz wave can be precisely tuned. Integrating multiple SWAs with distinct parameters as separate spatial channels on a single chip forms a gradient SWA. In this configuration, multiple discrete narrowband THz frequencies are generated simultaneously and collectively synthesize a long-duration, broadband THz pulse.

\begin{figure}[htbp] 
    \centering 
    \includegraphics[width=\textwidth]{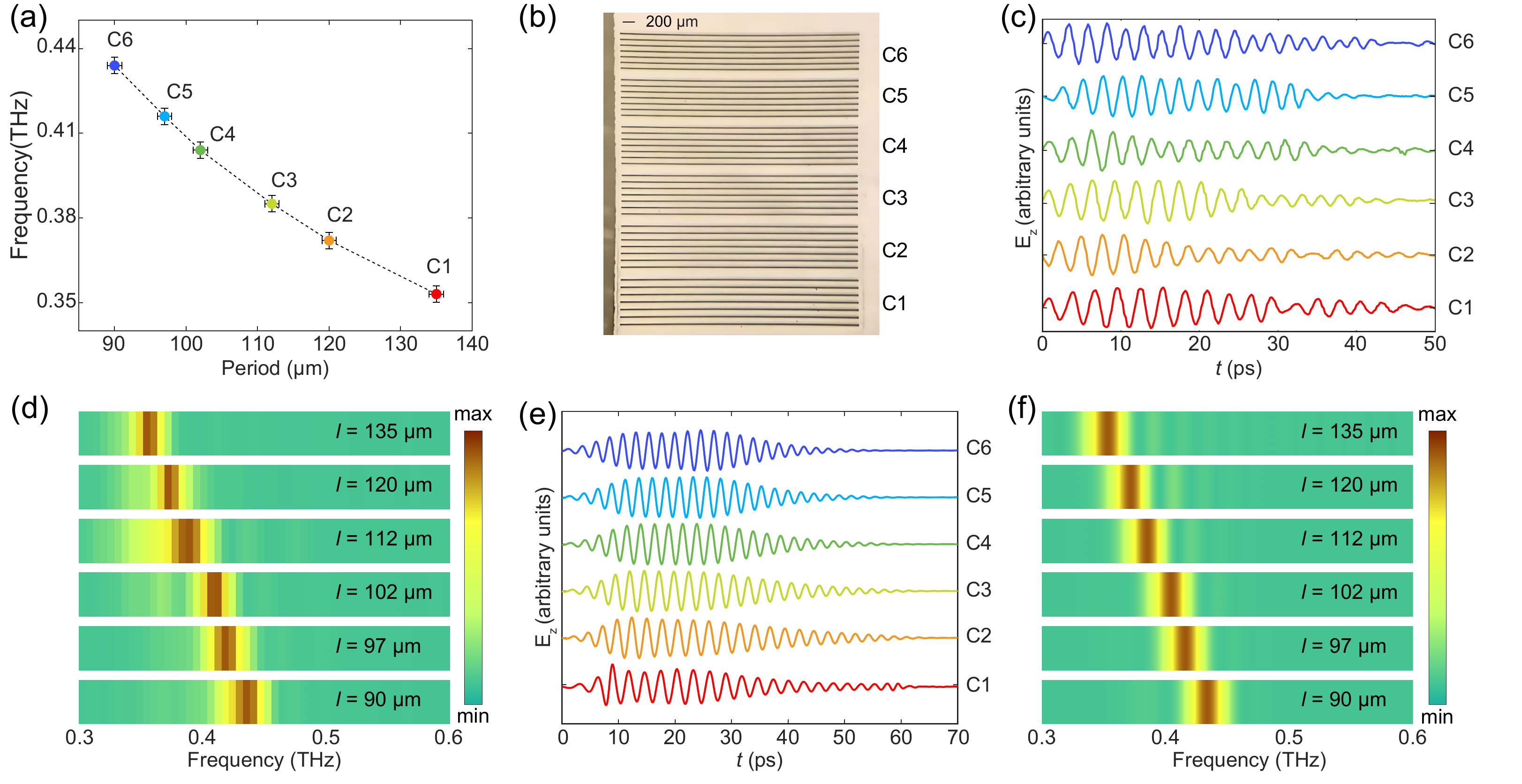}
    \caption{ The long-duration, broadband THz pulse generation in a gradient SWA. (a) Geometric parameters of the gradient SWA, and corresponding frequencies of the generated THz waves. (b) Appearance of the fabricated gradient SWA. (c) Time traces of the THz E-field generated in each channel of the gradient SWA. (d) Spectrum of all the THz pulses generated in the gradient SWA (experimental results). (e) Time traces of the THz E-field generated in each channel of the gradient SWA (simulation results). (f) Spectrum of all the THz pulses generated in the gradient SWA (simulation results).} 
    \label{FIG3} 
\end{figure}

To demonstrate this concept, a gradient SWA was fabricated on a 50‑$\mu$m‑thick $x$‑cut MgO:LN chip. The designed structure comprises 6 parallel channels, each consisting of 6 LN pillars and air slots to ensure efficient generation. The geometric parameters were optimized using Lumerical MODE Solutions, while all air slots have a fixed width of 8 $\mu$m, the LN pillar widths in successive channels are 135, 120, 112, 102, 97, and 90 $\mu$m, respectively.

As shown in Fig. 3(a), the designed gradient SWA is engineered to generate narrowband THz pulses at six discrete frequencies (0.353, 0.372, 0.385, 0.404, 0.416, and 0.434 THz) across its respective channels. An optical micrograph of the fabricated structure is presented in Fig. 3(b). Experimentally, THz pulses from each channel were detected individually via electro-optic sampling. The resulting temporal waveforms [Fig. 3(c)] show that each pulse lasts approximately 50 ps. Their corresponding spectral profiles, obtained by Fourier transformation, are plotted in Fig. 3(d), yielding center frequencies of 0.353, 0.370, 0.384, 0.406, 0.416, and 0.433 THz—closely matching the design targets.

Simulated temporal waveforms [Fig. 3(e)] exhibit pulse durations around 60 ps, and the simulated spectra [Fig. 3(f)] are in good agreement with experimental measurements. Minor discrepancies are attributed to fabrication tolerances. Collectively, Figs. 3(d) and 3(f) confirm that the spatially encoded channels simultaneously generate discrete THz frequencies, which together form a broadband spectrum. This result directly verifies the ability of the gradient SWA to synthesize long-duration pulses with wide spectral coverage.

The on-chip generation of broadband, long-duration THz pulses enables new modalities for light-matter interaction \cite{Shi2023-mi}, particularly in biosensing. Conventional THz sensing relies on time- or frequency-domain spectroscopy (THz-TDS \cite{koch2023terahertz} or THz-FDS \cite{liebermeister2021optoelectronic}), which require sequential scanning to acquire spectral data. In contrast, the channels of a gradient SWA can be engineered to emit discrete THz frequencies that directly match the fingerprint absorption peaks of target analytes. The presence of a specific substance can then be identified by detecting the absorption at its corresponding frequency, transforming the traditional scanning process into a parallel, instantaneous readout.

To demonstrate this scan-free sensing concept, we numerically simulate the detection of amino acid mixtures—histidine (His), tyrosine (Tyr), and glutamic acid (Glu)—using a specially designed gradient SWA. The permittivity of the amino acids is modeled using the Drude-Lorentz model \cite{YANG2025112139} (Supplementary Note 3), with characteristic absorption peaks \cite{choi2022chiral} located at 0.773, 0.963, and 1.201 THz, as shown in Fig. 4(a).

To target these three amino acids, we designed a gradient SWA comprising three independent channels, each engineered to generate a narrowband THz wave at a frequency matching the fingerprint absorption peak of one target analyte. The design is based on a 20‑$\mu$m‑thick LN slab with a uniform air‑slot width of 8 $\mu$m; the corresponding LN pillar widths are precisely tailored to 82.4 $\mu$m, 53.4 $\mu$m, and 38.2 $\mu$m in each channel to achieve the desired frequency‑specific generation.

\begin{figure}[htbp] 
    \centering 
    \includegraphics[width=0.8\linewidth]{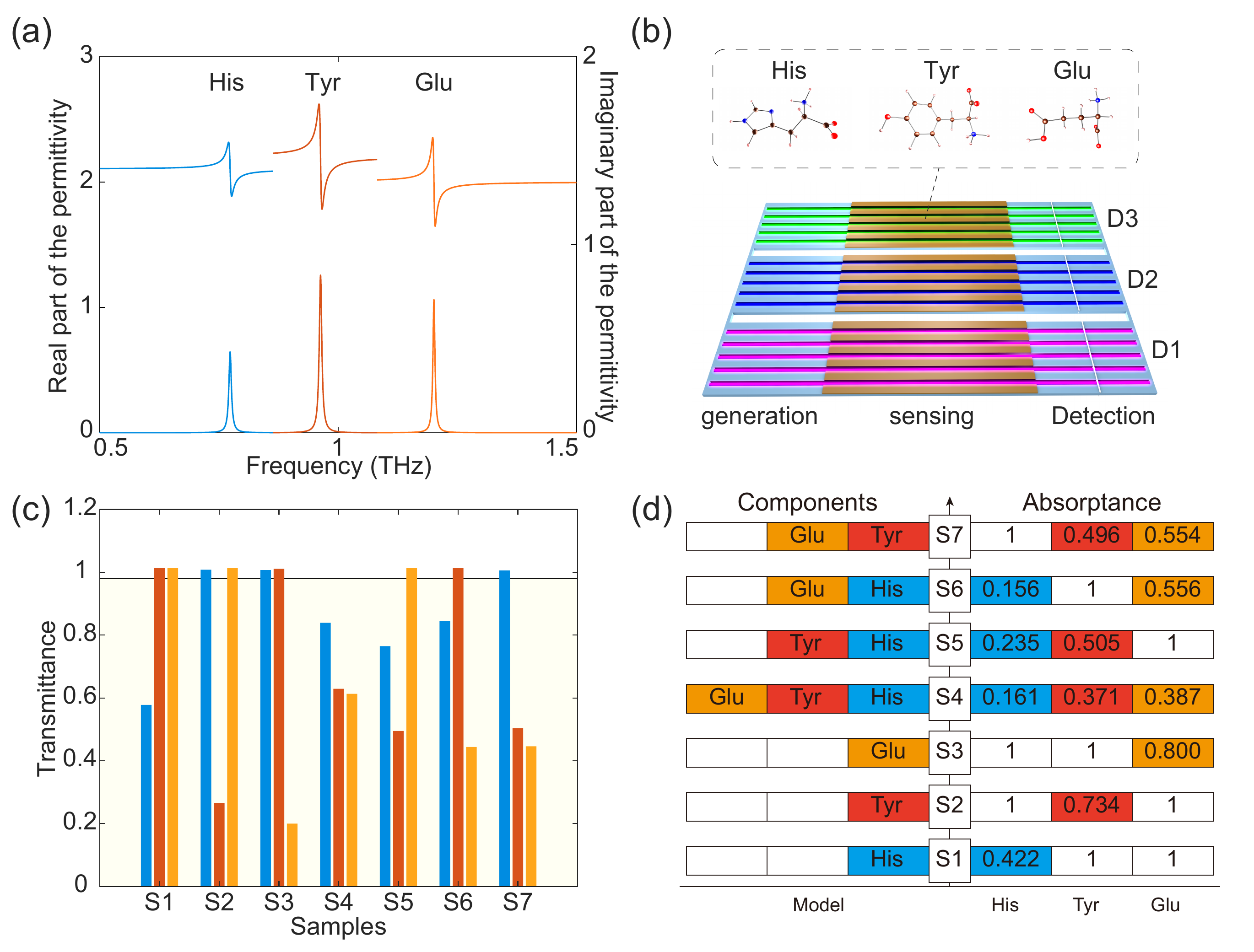} 
    \caption{ The on-chip scan-free identification of amino acids with the gradient SWA. (a), The calculated permittivity of His, Tyr and Glu. (b), Diagrams of detection of mixed amino acids. The mixture of amino acids, placed on the gradient SWA, is 5-$\mu$m thick and 2-mm-wide. The inset shows the structural formulas of His, Tyr and Glu. (c), The transmittance of detected samples. (d), Components detected in the samples according to the absorptance and the corresponding settings.} 
    \label{FIG4} 
\end{figure}

The simulation layout is shown in Fig. 4(b) and consists of three regions: THz generation, sample detection, and THz monitoring. In the generation region, the simultaneous emission of narrowband THz waves matching the absorption peaks is simulated (Supplementary Note 3). In the detection region, a 5‑$\mu$m‑thick, 2‑mm‑wide mixture of the amino acids, whose refractive index is modeled as a linear combination of the individual components, is placed on top of the gradient SWA. In the monitoring region, a frequency‑domain power monitor records the transmitted THz power behind the sample.

The transmittance $T(\omega_i)$ at a fingerprint absorption frequency $\omega_i$ is defined as the ratio of the transmitted power with the sample present to that without the sample:

\begin{equation}
\label{eq:T_ratio} %
T{(\omega_i)} = \frac{\int Re(P{(\omega_i)})^{\mathrm{mix}} \, \mathrm{d}S}{\int Re(P{(\omega_i)})^{\mathrm{ref}} \, \mathrm{d}S}
\end{equation}
where $Re(P{(\omega_i)})^{\mathrm{mix}}$ and $Re(P{(\omega_i)})^{\mathrm{ref}}$ denote the real part of the Poynting vector at frequency $\omega_i$ with and without the mixture in the detection region, respectively.

To systematically demonstrate the on‑chip scan‑free identification capability, we prepared a set of seven samples: the three individual amino acids, all three equimolar binary mixtures, and one equimolar ternary mixture. Figure 4(c) plots the simulated transmittance for all seven samples across the detection region, showing the distinct spectral signature of each mixture. Figure 4(d) compares the identified composition—derived from the fingerprint absorptance—with the predefined ground truth for each sample. The close match between the simulated results and the expected compositions confirms the successful implementation of scan‑free detection.
\section{conclusion}
In summary, we have realized and validated a robust on-chip platform that generates THz pulses with coexisting long duration and broad bandwidth—a capability previously inaccessible under the Fourier-transform limit. The core innovation lies in the spatial-spectral mapping of light across a gradient waveguide array, which confines different frequency components to diffraction-free channels and thereby overcomes the critical propagation bottleneck of free-space designs. This method not only enables scan-free identification of biomolecular mixtures, streamlining spectroscopic sensing, but also inaugurates a new architectural approach to integrated pulse generation. As a foundational building block, this platform paves the way for future integrated systems demanding precise spectral control, from high-speed THz communications to on-chip quantum information processing.

\section{acknowledgments} 
This work has been supported by the National Key Research and Development Program of China (2024YFA1409500), the National Natural Science Foundation of China (12574373, 12474344), the 111 Project (B23045).
\bibliography{myref}
\bibliographystyle{apsrev4-2}

\end{document}



\title{Supplementary Information for Spatial-spectral mapping for long-duration broadband terahertz pulse generation in on-chip waveguide arrays}


\author{Yibo Huang$^1$}
\author{Yao Lu$^1$}
\email[]{yaolu@nankai.edu.cn}
\author{Haoyu Duan$^1$}
\author{Chao Wang$^1$}
\author{Xitan Xu$^1$}
\author{Jiwei Qi$^1$}
\author{Qiang Wu$^{1, 2}$}
\email[]{wuqiang@nankai.edu.cn}
\author{Jingjun Xu$^{1, 2}$}
\email[]{jjxu@nankai.edu.cn}
\affiliation{$^1$ The Key Laboratory of Weak-Light Nonlinear Photonics, Ministry of Education, TEDA Applied Physics Institute and School of Physics, Nankai University, Tianjin 300457, China}
\affiliation{$^2$ Collaborative Innovation Center of Extreme Optics, Shanxi University, Taiyuan, Shanxi 030006, China}

\date{\today}


\maketitle

\section{Supplementary Note 1: Sample fabrication}
The waveguides utilized in our experiment are fabricated on 50 $\mu$m thick $x$-cut lithium niobite slabs using a chemically assisted femtosecond-laser direct writing technique. The experimental laser-machining setup for the sample fabrication is schematically depicted in Fig. S1. The machining laser beam has a central wavelength of 800nm, repetition rate of 1 kHz and pulse duration of 120 fs. The machining beam first passes through a shutter, which is used to block it when the sample is being moved from one machining region to the next. Then the beam passes through a power control system composed of a half-wave plate (HWP) and a Glan-Taylor prism (GTP), which is used to reduce and fix the power to 8 mW when machining. After passing through a dichroic mirror (DM), the machining beam is focused on the prepared slab mounted on an $xyz$-translation stage with computer-controlled actuators through an objective. To cut the slots out, the machining beam is focused on the slab at multiple depths spanning from the front to the back, which is realized by the translation stage with computer-controlled actuators. The machining process can be monitored in real time, which is realized by illuminating the sample using another beam of incoherent light and imaging on a CCD camera.
\begin{figure}[b] 
    \renewcommand{\thefigure}{S\arabic{figure}}
    \centering 
    \includegraphics[width=0.85\textwidth]{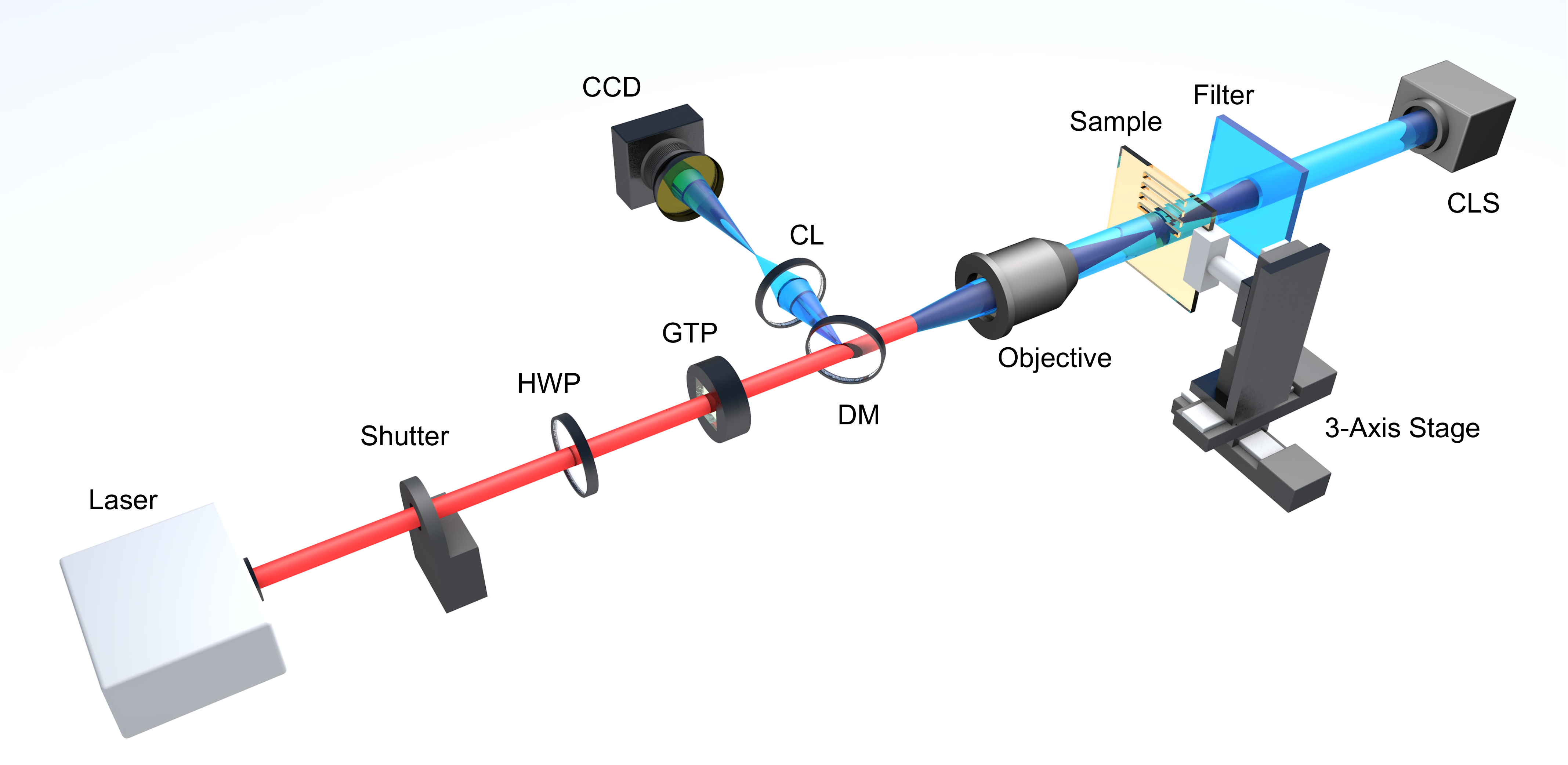}
    \caption{ Schematic diagram of the experimental laser-machining set-up for the sample fabrication. HWP: half wave plate, GTP: Glan Taylor prism, DM: dichroic mirror, CL: convex lens, CLS: cold light source.} 
    \label{SFIG1} 
\end{figure}
After the machining, the sample is immersed in hydrofluoric acid solution about 30 minutes to remove the silicon dioxide coating on the slab along with accumulated debris. Figure S2 shows a microscope image of the fabricated slot waveguide array (SWA). The SWA, i.e., the machining area, has a length of 4 mm, and the edge of the machining area and that of the plate are deliberately left at a certain distance for the stability of the sample. The width of all the slots is 8 $\mu$m, while the width of the LN pillars is 133 $\mu$m.

\begin{figure}[htbp] 
    \renewcommand{\thefigure}{S\arabic{figure}}
    \centering 
    \includegraphics[width=0.75\textwidth]{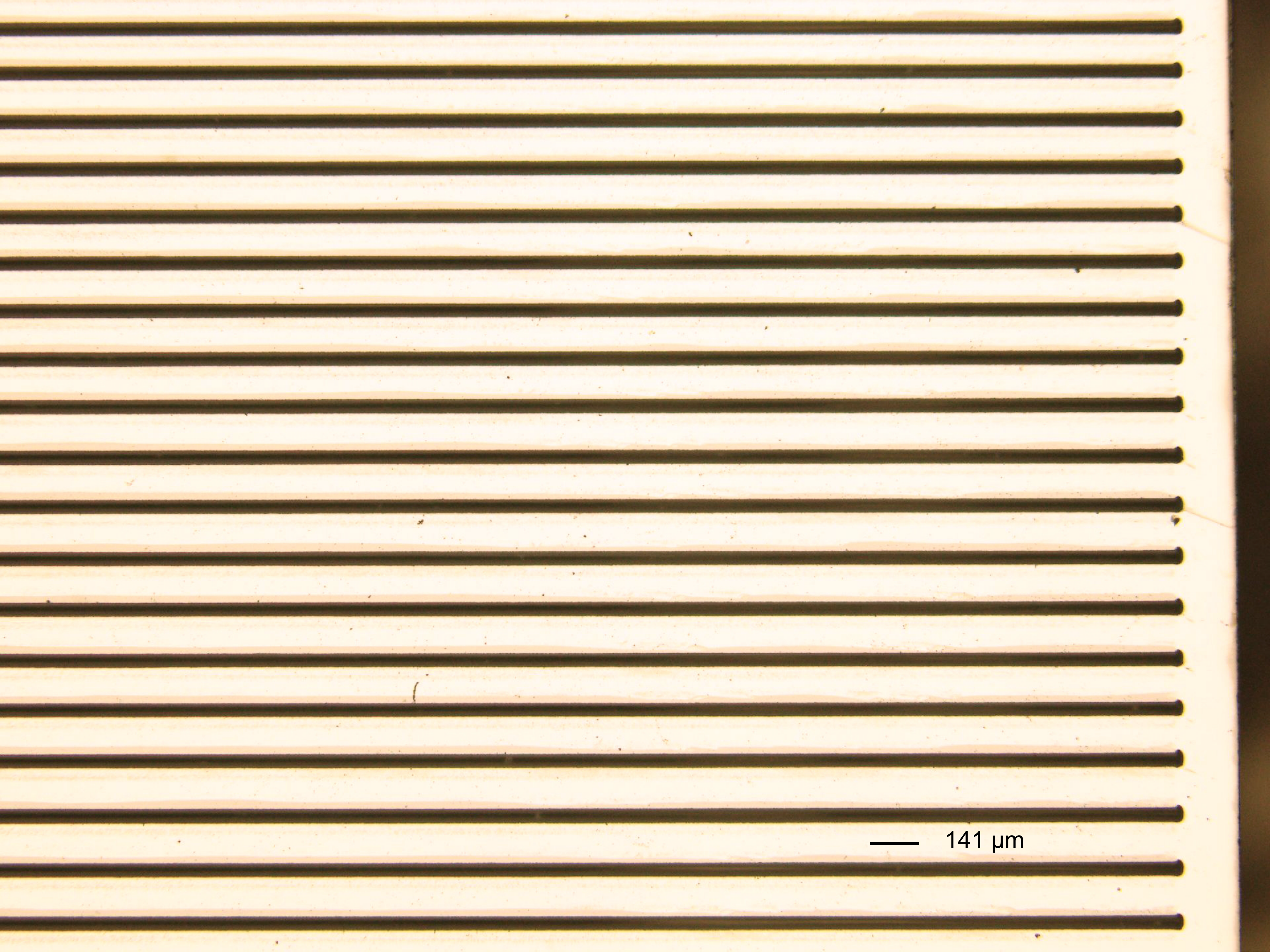}
    \caption{ Microscope image of the SWA used in the experiment. The width of all the slots and LN pillars are 8 and 133 $\mu$m, respectively.} 
    \label{SFIG2} 
\end{figure}

\section{Supplementary Note 2: THz generation and detection}
To generate and detect narrowband THz in the fabricated waveguides, we built a femtosecond laser-based electro-optical sampling system in the lateral excitation configuration, as dramatically shown in Fig. S3. Femtosecond laser pulses are split into two beams using a beam splitter, in which the pump beam is cylindrically focused into the waveguide to generate THz pulses, while the probe beam is delayed for detecting the generated THz waves. For the pump beam, a half-wave plate is used to control the polarization of it, making the polarization parallel to the LN optical axis. A chopper and a lock-in amplifier are used in combination to amplify detected signals. For the probe beam, a Glan-Taylor prism is used to ensure the polarization of it before incidence on the sample, and combined with a half-wave plate to control the power. Another unfabricated LN slab is rotated ninety degrees to compensate for the anisotropy of the sample, while a quarter-wave plate can convert the probe into circularly polarized light.
\begin{figure}[htbp] 
    \renewcommand{\thefigure}{S\arabic{figure}}
    \centering 
    \includegraphics[width=0.75\textwidth]{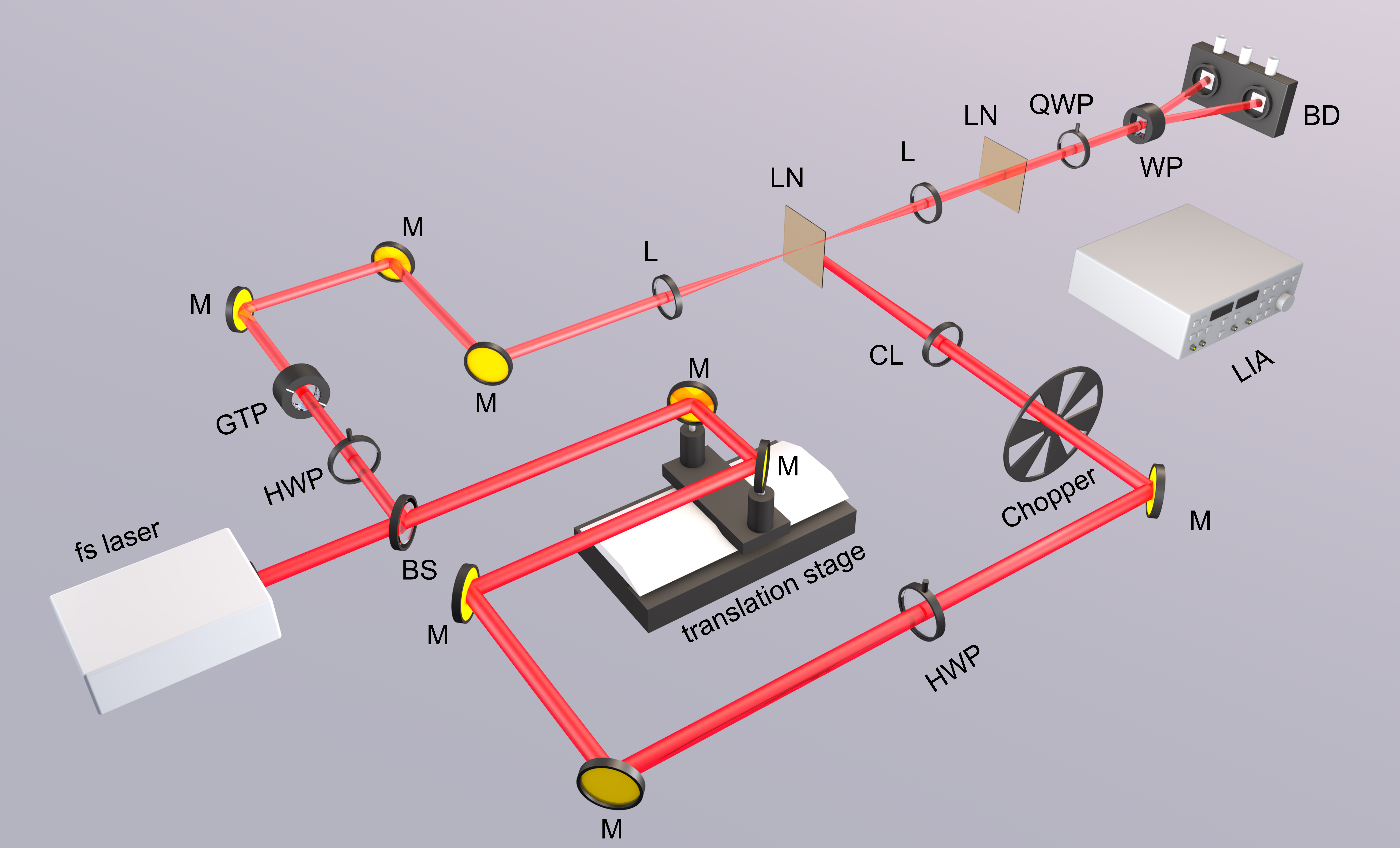}
    \caption{ Schematic diagram of the experimental electro-optical sampling system. HWP: half wave plate, GTP: Glan Taylor prism, CL: convex lens, M: mirror, L: lens, HWP: quarter wave plate, WP: Wollaston Prism, BD: balanced detectors, LIA: lock-in amplifier.} 
    \label{SFIG3} 
\end{figure}

\section{Supplementary Note 3: simulation setup}
The permittivity of the amino acids used in the simulation is described using the Drude-Lorentz model, which is as follows:

\begin{equation}
\varepsilon_i(\omega) = \varepsilon_{i\infty} + 
\frac{\mathrm{\Delta}\varepsilon_i \, \omega_{i0}^2}
{\omega_{i0}^2 - \omega^2 - 2\mathrm{i}\delta_i\omega}
\end{equation}
where $\varepsilon_{i\infty}$ is the high frequency constant term, $\omega_{i0}$ is the resonance frequency, ${\mathrm{\Delta}\varepsilon_i}$ is the change in relative permittivity, and $\delta_i$ is the damping parameter. The resonance frequencies of His, Tyr and Glu are 
$\omega_1 = 2\pi \times 0.773$, 
$\omega_2 = 2\pi \times 0.963$ and 
$\omega_3 = 2\pi \times 1.201$~THz, respectively. 
In the simulation, other parameters are set as: 
$\varepsilon_{1\infty} = 2.1$, 
$\varepsilon_{2\infty} = 2.2$, 
$\varepsilon_{3\infty} = 2$, 
$\Delta\varepsilon_1 = 0.00354$, 
$\Delta\varepsilon_2 = 0.00554$, 
$\Delta\varepsilon_3 = 0.003$, 
$\delta_1 = 2\pi \times 0.02$~THz, 
$\delta_2 = 2\pi \times 0.02$~THz, 
$\delta_3 = 2\pi \times 0.016$~THz.

Targeting these three amino acids, a gradient SWA with three channels is designed, in which channels are engineered to generate narrowband THz waves at frequencies of 0.773, 0.963 and 1.201 THz, respectively. In the simulations, the THz generation region is set to 4 mm in width. The time traces of the THz E-field generated in each channel is shown in Fig. S4(a), lasting around 60 ps. Spectrum of all the generated THz pulses in simulations are obtained by applying Fourier transformation to the time-domain signals, as shown in Fig. S4(b).
\begin{figure}[htbp] 
    \renewcommand{\thefigure}{S\arabic{figure}}
    \centering 
    \includegraphics[width=0.95\textwidth]{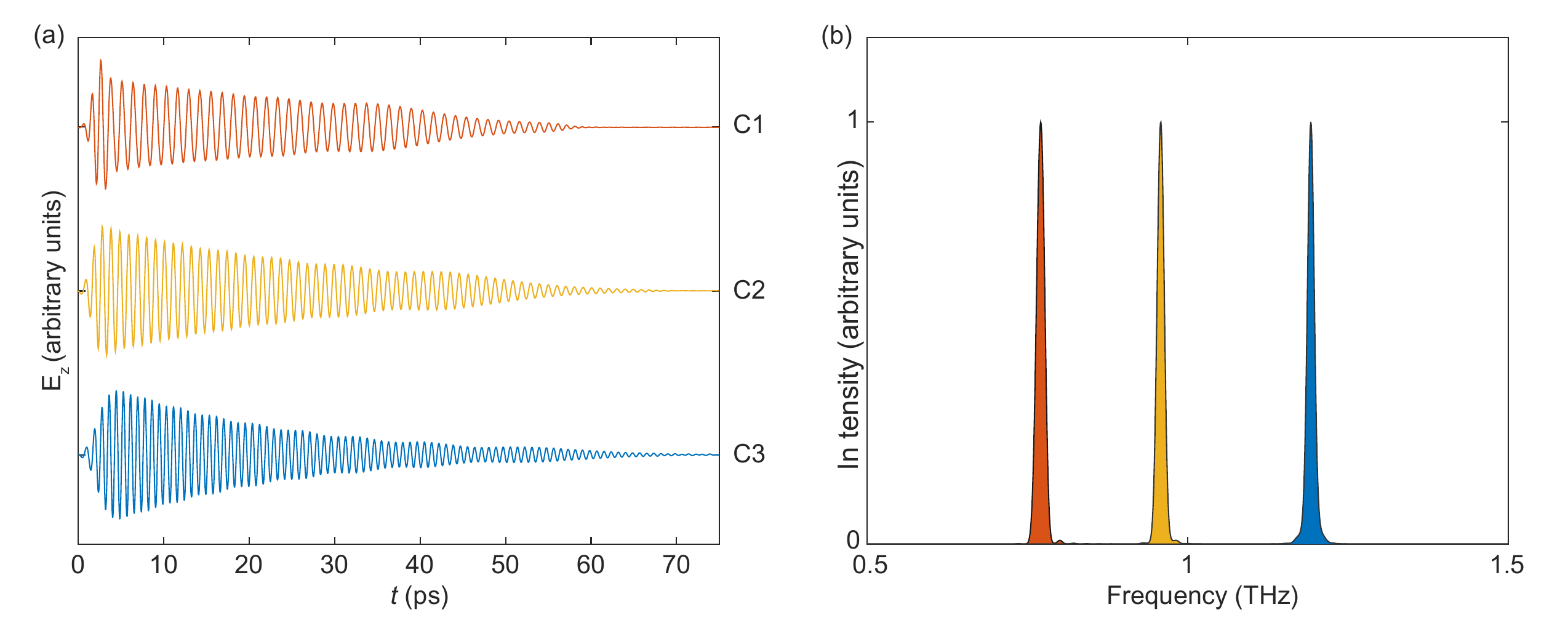}
    \caption{ Narrowband THz waves matching the fingerprint absorption peaks. (a) Time traces of the THz E-field generated in each channel of the designed gradient SWA. (b) Spectrum of all the THz pulses generated in the designed gradient SWA.} 
    \label{SFIG4} 
\end{figure}